\documentstyle[11pt,newpasp,twoside,epsfig]{article}
\def\edcomment#1{\iffalse\marginpar{\raggedright\sl#1\/}\else\relax\fi}
\reversemarginpar

\begin{document}
\title{V382 Vel: a ``shocking'' supersoft x-ray source? 
}
 \author{Marina Orio}
\affil{Turin Astronomical Observatory, I-10025 Pino Torinese (TO), Italy, 
and Astronomy Department,  470 N. Charter Str., 53706 Madison,
WI, USA}
\author{Arvind Parmar}
\affil{ Astrophysics Division, Space Science Dept. of ESA, Postbus 299,
2200 AG Noordwijk, The Netherlands}

\begin{abstract}

 Nova Vel 1999 (V382 Vel) was observed with BeppoSAX twice, 15 days and 6
months after the optical maximum. A hard X-ray source was detected
in the first observation, while the second time
also a very luminous supersoft X-ray source was detected.
 The continuum observed  
in the supersoft range with the BeppoSAX LECS  cannot be 
fitted with atmospheric models of hot hydrogen burning
while dwarfs. We suggest that we are observing
instead mainly a ``pseudocontinuum'', namely a blend of
very strong emission lines in the supersoft X-ray range.
\end{abstract}

\section{Introduction}

Nova Velorum 1999 (V382 Vel) was the second brightest nova of this
 half of the century (V=2.6) (Seargent \& Pearce, 1999).
It  was a Ne-O-Mg nova, and quite a ``fast'' one, with v$\simeq$4000 Km/s,
 t$_2$= 6 d, t$_3$=10 d (Della Valle et al. 1999, Shore
et al. 1999).

Two mechanisms of X-ray emission are known for classical
 novae in outburst: they teach us about the mass ejection process 
and about the final outcome of these systems (see Kahabka et al. 1999
and references therein).
{\it ``Hard''} X-ray emission (thermal bremstrahlung
continuum at plasma temperatures in the range 0.5-10 KeV, and possibly 
additional emission lines) with luminosities 10$^{33-34}$ erg s$^{-1}$
has been observed and attributed to 
{\it shocks}. The nova outburst is normally due to a radiation pressure driven
wind and not to a shock wave, however 
shocks would be produced in interacting winds or interaction
between the ejecta and the circumstellar medium.
Luminous {\it ``supersoft''} X-ray emission (luminosity
of the order 10$^{27-38}$ erg s$^{-1}$) has also been
observed and attributed instead 
to residual hydrogen burning in a shell on the white
dwarf remnant. We expect to detect in this case an 
atmospheric continuum at T$_{eff}$=20-80 eV 
and absorption edges of the white dwarf (or {\it emission
edges} if the effective temperature is very high).

\section{The BeppoSAX observations}

V382 Vel was observed for the
first time with BeppoSAX 15 days after the optical maximum, on 1999 June
7-8 for 42.5 Ks with the two MECS, for 13.5 Ks with the LECS, for
23.3 Ks with the PDS. No other classical nova had ever
been observed immediately after outburst and in such a broad
energy range.  The nova  was then observed a second time on 1999 November
23 for 25.9 Ks with the MECS,  12.4 Ks with the LECS, 12.4
Ks with the PDS.  In the June 1999 observation
 the nova was detected with a count rate 0.1537$\pm$0.0020 cts s$^{-1}$
and 0.0620$\pm$0.0026 cts s $^{-1}$ in the MECS and LECS,
respectively (Orio et al. 1999a).  At higher energies the 2$\sigma$ upper
limits obtained with the PDS were 0.048 cts s$^{-1}$ in the   15-30 keV  
band and 0.074 cts s$^{-1}$ in the 15-60 keV band.
The results  are described in detail
in Orio et al. 2000a and b. The spectrum was characterized by a
a very large absorption of the ejecta, which also shielded the central
source.The best fit to the data is obtained with a 
a {\sc vmekal} model of thermal plasma with kT=6.2 keV, flux F$_x$=1.8
10$^{-11}$ erg cm$^{-2}$ s$^{-1}$ (corresponding to
an {\it unabsorbed} flux 4.3 $\times$ 10$^{-11}$
erg cm$^{-2}$ s$^{-1}$), N(H)=1.7 10$^{23}$ cm$^{-2}$,
and reduced iron abundance (Orio et al. 2000b).
\begin{figure}
\hspace{3 true cm} \psfig{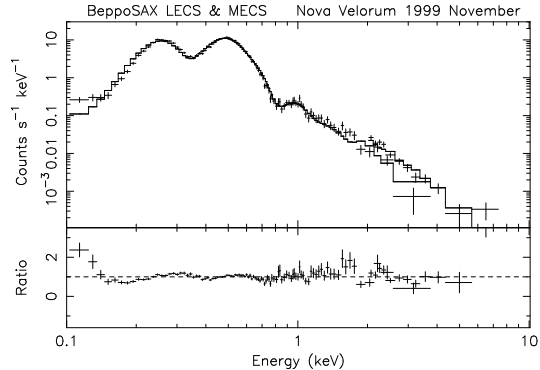}
\caption{Spectra observed in the 0.1-10 keV range 
with the BeppoSAX LECS and MECS in November 1999
 and best fit with a model atmosphere studied by
 Hartmann \& Heise (1997).  The fit is not acceptable in the supersoft
range where most of the flux is detected. The lower 
panel shows the residuals in counts per energy bin.}
\end{figure}
\section{The second BeppoSAX observation}

In the second BeppoSAX observation, which we want to discuss more in detail
in this paper, the count rate measured with the
LECS was extremely high, 3.468$\pm$0.003 cts s$^{-1}$,
due to the emergence of the supersoft X-ray source
 (Orio et al.  1999b). The MECS count rate had instead decreased
to 0.0449$\pm$0.0015 cts s$^{-1}$ (more than a factor 3 lower than in June).
Again, there was no PDS detection  with 2$\sigma$ upper limit was
0.080 cts s$^{-1}$ in the 15-50 keV range. The portion of the 
LECS spectrum {\it above} 0.8 keV can be fitted simultaneously
with the MECS spectrum with a {\sc mekal} model of thermal plasma
with parameters:  N(H)$\simeq$ 2 $\times$ 10$^{21}$
cm$^{-2}$, kT$\simeq$700 eV, unabsorbed flux $\simeq$ 10$^{-12}$ erg
cm$^{-2}$ s$^{-1}$ (the reduced $\chi^2$ is 1.5).
 We tried to fit instead the LECS spectrum {\it below} 0.8 keV
(where most of the flux is detected)
with {\it a)} a blackbody, {\it b)} a blackbody with absorption
edges, and {\it c)} with a more detailed model atmosphere
 (Hartmann \& Heise 1997).
1999 and references therein). We found that 
a reasonable fit cannot be obtained with
of these models.  In Fig. 1 we show as an example
the fit with {\it c)},  a  model atmosphere studied by
 Hartmann \& Heise (1997), with log(g)=8.0, blackbody temperature
47 eV, and kT=0.9 keV for  a bremstrahlung component at higher energy. 
This fit is not acceptable.

In Fig. 2 we fitted instead the whole LECS and MECS spectrum, from
0.1 to 10 keV, with continuum+lines. Only narrow emission lines, 
that can be produced in the nebula but not in the
white dwarf atmosphere, make the fit possible. 
  We obtain a much more acceptable
 fit than with any atmospheric model (reduced $\chi^2$=1.9). In this example
we obtained N(H)=2 $\times 10^{21}$ cm$^{-2}$ (very close
to the interstellar value). To explore
 some possible ``complexity'', we
assumed that the continuum is due to the Wien tail of a
a blackbody-like component with kT$\simeq$20 eV, to a  bremsstrahlung component
at kT=47 eV and an additional powerlaw component with photon index 4.9.
The following, narrow emission lines were overimposed:
N VII at 500.36 eV,  O VII at 561.04 keV, Ne X at 1.022 keV,
Si XV at 2.01 keV,  FeXV at 6.67 keV. The acceptable fit
is obtained by adding the emission lines and the quality of
the result is not very dependent on the continuum assumed in the
supersoft range.
The lines of N VII, O VII, and Ne X were indeed observed shortly later
with the CHANDRA LETG (on December 30 2000: courtesy of S. Starrfield et al.,
2000). It is our understanding from preliminary results, however,
that the continuum detected with the LETG and attributed to
the central source   was not sufficiently
hot to produce the lines by photoionization.
We hypothesize therefore that the emission lines were produced by shocks. 

\begin{figure}
\hspace{3 true cm} \psfig{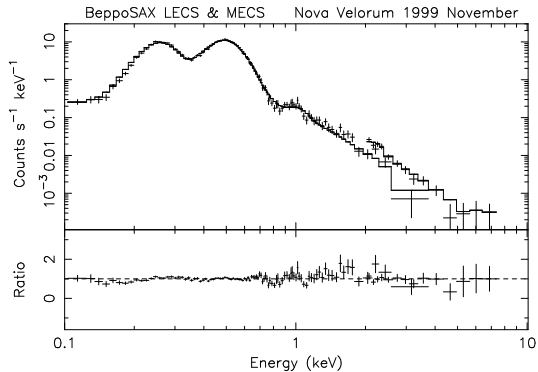}
\caption{Here we fitted the LECS and MECS spectra, observed in November 1999 
in the 
0.1 to 10 keV range, with continuum and emission lines in
the supersoft range (see text).}
\end{figure}
\section{Do supersoft X-ray observations of novae need to be revisited?}

 A low effective temperature ($\leq$ 20 eV)
of the the post-nova white dwarf atmosphere might explain why
 a ionization nebula
was been detected only in H$\alpha$ for N Cyg 1992 (Casalegno et al.
2000) while {\it other ionization lines} (indicating
a higher ionization potential) {\it were not present} in the nebula. 
We speculate therefore that also N Cyg 1992 (Krautter et al.
1996, Balman et al. 1998) might have been {\it cooler} than
it appeared fitting the ROSAT PSPC spectrum. We remind that
we came to the proposed possible explanation of 
the V382 Vel X-ray spectrum thanks to information on the CHANDRA data,
 obtained 6 weeks later by Starrfield et al. (2000). Both BeppoSAX
and the ROSAT PSPC in fact could only detect a pseudo-continuum but 
{\it cannot resolve} the high ionization lines.

The difficulty in explaining the BeppoSAX spectrum
without invoking the never-before predicted emission lines
in the supersoft range 
(indeed observed shortly afterwards with CHANDRA),
suggests that the observed ``supersoft X-ray source'' in
V382 Vel is mostly due to non-resolved lines. The LECS must have detected a
``pseudocontinuum'' due to  a blend
of high ionization emission lines.
 
 Very strong nebular emission lines in the supersoft
X-ray energy range may indicate interesting possibilities.
If ``cool'' continua with strong emission
lines overimposed in the supersoft range are detected also for other
novae (observed with Chandra or XMM in the future) we face new questions. Is
it common to have powerful shocks in the nova wind even many months
after the outburst? Should we consider
instead a line driven wind in addition to
the radiation driven wind to explain the nova mass loss? The nova theory is
going to become more detailed and refined once the X-ray spectrum is known in 
detail for a statistically meaningful sample of objects.
 If the supersoft X-ray source
in post-novae is generally ``cooler'' than it was believed,
the fact that ROSAT detected very few post-nova-outburst white dwarfs 
is also explained (Orio et al. 2000c). The effective temperature must
have been below the detectable range with the ROSAT PSPC.

\end{document}